\documentclass[aps,prl,twocolumn,superscriptaddress,preprintnumbers]{revtex4-1}

\usepackage{graphicx}
\usepackage{lineno} 
\usepackage{amsmath}
\usepackage{amssymb}
\usepackage{etoolbox} 
\usepackage{bm}
\usepackage{epsfig}
\usepackage{tikz}
\usepackage{array}
\usepackage{xcolor}
\usepackage{soul}
\usepackage{colortbl}
\usepackage{cellspace}
\definecolor{Myyellow}{HTML}{FFFF00}
\definecolor{Myorange}{HTML}{FFCE00}

\newcommand*\linenomathpatch[1]{%
  \cspreto{#1}{\linenomath}%
  \cspreto{#1*}{\linenomath}%
  \csappto{end#1}{\endlinenomath}%
  \csappto{end#1*}{\endlinenomath}%
}

\linenomathpatch{equation}
\linenomathpatch{gather}
\linenomathpatch{multline}
\linenomathpatch{align}
\linenomathpatch{alignat}
\linenomathpatch{flalign}
\def\beqn{\begin{eqnarray}}
\def\eeqn{\end{eqnarray}}

\begin{document}

\title{How well does nonrelativistic QCD factorization work at next-to-leading order?}

\author{Nora~Brambilla}
\affiliation{Technical University of Munich, TUM School of Natural Sciences,
Physics Department,\\
James-Franck-Strasse 1, 85748 Garching, Germany} 
\affiliation{Technical University of Munich, Institute for Advanced Study,\\ Lichtenbergstrasse 2a,
85748 Garching, Germany} 
\affiliation{Technical University of Munich, Munich Data Science Institute, \\ Walther-von-Dyck-Strasse 10, 85748
Garching, Germany} 
\author{Mathias~Butenschoen}
\affiliation{II. Institut f\"ur Theoretische Physik, Universit\"at Hamburg,\\
Luruper Chaussee 149, 22761 Hamburg, Germany} 
\author{Xiang-Peng~Wang}
\affiliation{Institute of Particle Physics and Key Laboratory of Quark and Lepton Physics (MOE), Central
China Normal University, Wuhan, Hubei 430079, China} 
\affiliation{Technical University of Munich, TUM School of Natural Sciences,
Physics Department,\\
James-Franck-Strasse 1, 85748 Garching, Germany} 

\date{\today}

\begin{abstract}
We perform a thorough investigation of the universality of the long distance matrix elements (LDMEs) of nonrelativistic QCD factorization based on a next-to-leading order (NLO) fit of $J/\psi$ color octet (CO) LDMEs to high transverse momentum $p_T$ $J/\psi$ and $\eta_c$ production data at the LHC. We thereby apply a novel fit-and-predict procedure to systematically take into account scale variations, and predict various observables never studied in this context before. In particular, the LDMEs can well describe $J/\psi$ hadroproduction up to the highest measured values of $p_T$, as well as $\Upsilon(nS)$ production via potential NRQCD based relations. Furthermore, $J/\psi$ production in $\gamma \gamma$ and $\gamma p$ collisions is surprisingly reproduced down to $p_T=1$~GeV, as long as the region of large inelasticity $z$ is excluded, which may be of significance in future quarkonium studies, in particular at the EIC and the high-luminosity LHC. In addition, our summary reveals an interesting pattern as to which observables still evade a consistent description.
\end{abstract}

\maketitle

\paragraph{Introduction and overview ---} The hierarchy of energy scales $m_Q\gg m_Qv\gg m_Qv^2$ makes heavy quarkonium production an ideal laboratory to study both the perturbative and nonperturbative aspects of QCD. 
Here, $m_Q$ stands for the heavy quark mass and $v$ for the relative velocity of the heavy quark and antiquark in the rest frame of the quarkonium. 
The most prominent approach to describe quarkonium production and decay is via non-relativistic QCD (NRQCD)~\cite{Bodwin:1994jh}.
This is an effective field theory that comes with a conjectured factorization formula,
according to which a quarkonium, $H$, production cross section factorizes into perturbatively calculable short-distance coefficients (SDCs) and universal non-perturbative long-distance matrix elements (LDMEs) $\langle \mathcal{O}^H(n)\rangle$, 
where $n={^{2S+1}L}_J^{[1/8]}$ denotes an intermediate color singlet ($^{[1]}$, CS) or octet ($^{[8]}$, CO) heavy quark-antiquark state with spin $S$, orbital angular momentum $L$ and total angular momentum $J$. 
NRQCD predicts the LDMEs to scale with certain powers of $v^2$ via velocity scaling rules~\cite{Lepage:1992tx}. 
These rules predict the leading (next-to-leading) $v^2$ contributions to stem for $\psi(nS)$ and $\Upsilon(nS)$ production from $n={^3S}_1^{[1]}$ (${^1S}_0^{[8]}$, ${^3S}_1^{[8]}$, ${^3P}_J^{[8]}$) and for $\eta_c$ production from $n={^1S}_0^{[1]}$ (${^3S}_1^{[8]}$, ${^1S}_0^{[8]}$, ${^1P}_1^{[8]}$). 
The LDMEs are, however, not all independent. 
Up to corrections of order $v^2$, the LDMEs of $J/\psi$ and $\eta_c$ are mutually related via heavy quark spin symmetry (HQSS)~\cite{Bodwin:1994jh}, and the various $\psi(nS)$ and $\Upsilon(nS)$ LDMEs by relations~\cite{Brambilla:2022rjd,Brambilla:2022ayc} based on potential NRQCD (pNRQCD)~\cite{Pineda:1997bj,Brambilla:1999xf,Brambilla:2004jw,Brambilla:2020ojz,Brambilla:2021abf}.

To scrutinize the LDME universality is an ongoing task. 
Currently, these tests are performed at next-to-leading order (NLO) in the strong coupling constant $\alpha_s$, and may, for LDMEs concerning $J/\psi$, be summarized as follows: 
The LDMEs of the global fit \cite{Butenschoen:2011yh} to 194 points of $J/\psi$ production data in $pp$, $p\overline{p}$, $\gamma p$, $\gamma\gamma$ and $e^+e^-$ scattering describe the fitted observables reasonably well, but result in discrepancies for $\eta_c$~\cite{Butenschoen:2014dra} and $J/\psi+W$ or $Z$~\cite{Butenschoen:2022wld} yields at the LHC, and predict strong transverse $J/\psi$ polarization, not observed at Tevatron or the LHC~\cite{Butenschoen:2012px}. 
All other LDME fits are restricted to high transverse momentum $p_T$ hadroproduction data, with low-$p_T$ cuts of typically 7~GeV or higher~\cite{Ma:2010yw,Gong:2012ug,Han:2014jya,Zhang:2014ybe,Bodwin:2015iua,Brambilla:2022rjd,Brambilla:2022ayc}. 
They are successful at describing the fitted high $p_T$ yields and the measured $J/\psi$ polarization, but describe neither the hadroproduction data at lower $p_T$ 
nor the inelasticity variable $z$ integrated photoproduction~\cite{Butenschoen:2012qr}, and the $J/\psi+W$ or $Z$ data~\cite{ATLAS:2014yjd,ATLAS:2014ofp,ATLAS:2019jzd} at best marginally~\cite{Butenschoen:2022wld}. 
$\eta_c$ hadroproduction is almost exclusively determined by $\langle{\cal O}^{J/\psi}({^1S}_0^{[8]})\rangle$. 
The fit results of Refs.~\cite{Ma:2010yw,Brambilla:2022rjd,Brambilla:2022ayc} leave $\langle{\cal O}^{J/\psi}({^1S}_0^{[8]})\rangle$ poorly constrained, predicting $\eta_c$ production therefore only with very large uncertainties. 
On the other hand, Refs.~\cite{Gong:2012ug,Bodwin:2015iua} lead to an overshoot of $\eta_c$ data. Only analyses~\cite{Han:2014jya,Zhang:2014ybe} describe $\eta_c$ production well, 
because they use LHCb $\eta_c$ data directly as input to constrain $\langle{\cal O}^{J/\psi}({^1S}_0^{[8]})\rangle$. 
The latter approach thus appears to be the most promising fit strategy, and is the one we adopt here.

In our NLO analysis, we fit the three $J/\psi$ CO LDMEs to prompt LHCb $\eta_c$~\cite{LHCb:2014oii,LHCb:2024ydi} and CMS $J/\psi$~\cite{CMS:2015lbl} production yields, the latter data ranging from $p_T=10$~GeV to 120~GeV. 
After showing that these LDMEs lead to a good description of the CMS $J/\psi$ polarization data~\cite{CMS:2015lbl}, we go on to make predictions for observables never studied in this context before: 
We show that our LDMEs can predict well recent ATLAS $J/\psi$ hadroproduction data with $p_T$ ranging up to 360~GeV~\cite{ATLAS:2023qnh}. 
We further show that, contrary to common perceptions, these fits can well describe $\gamma\gamma$ scattering at LEP~\cite{DELPHI:2003hen} and $J/\psi$ photoproduction at HERA~\cite{H1:2010udv,ZEUS:2012qog} even at $p_T$ as low as 1~GeV, as long as we consider the region $z<0.6$.
Even $\Upsilon(nS)$ production~\cite{ATLAS:2012lmu} is well reproduced, 
while $J/\psi+Z$ production~\cite{ATLAS:2014ofp} remains  intricate to be interpreted. 
All these findings are nontrivial tests of NRQCD factorization never performed before.

A technological advancement is our consistent fit-and-predict procedure which systematically takes into account the effect of scale variations, the largest source of SDC uncertainty.
In most NRQCD analyses, e.g. Refs.~\cite{Ma:2010yw,Gong:2012ug,Han:2014jya,Zhang:2014ybe}, scale variations are completely ignored. 
In Refs.~\cite{Butenschoen:2011yh,Butenschoen:2014dra,Butenschoen:2022wld,Butenschoen:2012px,Butenschoen:2012qr,Butenschoen:2022qka,Butenschoen:2012qr,Butenschoen:2011ks,Butenschoen:2010rq,Butenschoen:2019npa,Butenschoen:2013pxa}, scale uncertainties are considered in predictions and plots, but, unrealistically, assumed to be uncorrelated to the LDME errors. 
In analyses~\cite{Bodwin:2015iua,Brambilla:2022rjd,Brambilla:2022ayc}, scale uncertainties are treated as global theory errors in the fits, unable, however,
to predict scale uncertainties of other observables with own individual scale dependencies. 
We therefore proceed as follows. 
We do three separate fits, with the renormalization and factorization scales $\mu_r$ and $\mu_f$ set equal and equal to $\mu_0/2$, $\mu_0$ and $2\mu_0$, respectively, with $\mu_0$ our default hard scale. 
For prediction plots, we then create three error bands, each created using the LDME uncertainties of one fit, in combination with the SDCs evaluated with the respective scale choice. 
Our overall uncertainty is then the envelope of the three error bands created this way.

\paragraph{Details and input of the calculation ---} We calculate the SDCs needed for hadroproduction yields using the dipole subtraction based codes described in Refs.~\cite{Butenschoen:2019lef,Butenschoen:2020mzi,Butenschoen:2022wld}. 
For $J/\psi$ polarization and for $J/\psi$ production in $\gamma p$ and $\gamma\gamma$ scattering, we calculate the SDCs using the phase-space slicing~\cite{Harris:2001sx} based codes described in Refs.~\cite{Butenschoen:2012px,Butenschoen:2009zy,Butenschoen:2011yh,Butenschoen:2020mzi}. 
We use three (four) light quark flavours for charmonia (bottomonia) in loops and external legs, heavy quarks only for the outgoing $Q\overline{Q}$ pair.
We set the default hard scale to be $\mu_0=m_T=(p_T^2 +4m_Q^2)^{1/2}$, the quarkonium transverse mass, and the NRQCD scale to be $\mu_{\Lambda}=m_Q$. 
We choose the on-shell charm and bottom masses to be $m_c=1.5$ GeV and $m_b=4.75$ GeV. 
As proton parton distribution function (PDF) set, we use CTEQ6M~\cite{Pumplin:2002vw}, and for $\alpha_s(\mu_r)$ correspondingly the two loop running formula with the asymptotic scale parameter $\Lambda_{\rm QCD}^{(n_f=4)} = 326$~MeV ($\Lambda_{\rm QCD}^{(n_f=5)} = 226$~MeV) for charmonia (bottomonia). 
In $\gamma p$ and $\gamma\gamma$ scattering, we use the Weizs\"acker--Williams approximation for the photon flux and for resolved photons the AFG04\_BF PDF set~\cite{Aurenche:2005da}. 
We evaluate $\psi(2S)\to J/\psi$ feeddown assuming $p_{T,\psi(2S)}/p_{T,J/\psi}=m_{\psi(2S)}/m_{J/\psi}=1.19$ \cite{ParticleDataGroup:2024cfk}. 
The $\chi_{cJ}\to J/\psi$ feeddown contributions are estimated for $J/\psi$ hadroproduction  by interpolating ATLAS data~\cite{ATLAS:2014ala}, and for $J/\psi +Z$ production to be $20\%$ of the direct $^3S_1^{[8]}$ production channel, based on Ref.~\cite{Butenschoen:2022wld} and the fitted value of $\langle\mathcal{O}^{\chi_{c0}}(^3S_1^{[8]})\rangle$ in Ref. \cite{Ma:2010vd}. 
We neglect $\chi_{cJ}$ feeddowns for our polarization predictions due to their small influence on $\lambda_{\theta}$~\cite{Brambilla:2022ayc}, 
and in $\gamma p$ and $\gamma\gamma$ scattering due to their smallness there~\cite{DELPHI:2003hen,H1:2010udv,ZEUS:2012qog}. 
$\chi_{bJ}\to \Upsilon(nS)$ feeddowns are estimated by using the LHCb measured feeddown fractions~\cite{LHCb:2014ngh}. 
$h_c\to\eta_c$ feeddown is again negligible~\cite{Butenschoen:2014dra} and neglected, too. 
We adopt the values for all branching fractions from Ref.~\cite{ParticleDataGroup:2024cfk}. 
Our fits are standard least-square fits of the three parameters $\langle\mathcal{O}^{J/\psi}(^3S_1^{[8]})\rangle$,  $\langle\mathcal{O}^{J/\psi}(^1S_0^{[8]})\rangle$ and $\langle\mathcal{O}^{J/\psi}(^3P_0^{[8]})\rangle/m_c^2$, in this order, 
taking into account experimental errors and a global 30\% theory error due to unknown relativistic corrections. 
We relate our fit parameters to the three $\eta_c$ CO LDMEs via HQSS relations~\cite{Bodwin:1994jh}, 
and to the $\psi(2S)$ and $\Upsilon(nS)$ CO LDMEs according to the pNRQCD derived relations~(3.47)--(3.48) of Ref.~\cite{Brambilla:2022ayc}, where, however, instead of (3.47b) we use its $1/\beta_0$ expanded form up to $O(1/\beta_0)$.
The CS LDMEs are not fitted, but fixed as $\langle \mathcal{O}^{J/\psi}(^3S_1^{[1]})\rangle = 1.16$ GeV$^3$~\cite{Eichten:1995ch}, $\langle \mathcal{O}^{\psi(2S)}(^3S_1^{[1]})\rangle = 0.76$ GeV$^3$~\cite{Eichten:1995ch}, $\langle \mathcal{O}^{\Upsilon(3S)}(^3S_1^{[1]})\rangle =3.54$ GeV$^3$~\cite{Eichten:1995ch} and $\langle \mathcal{O}^{\eta_c}(^1S_0^{[1]})\rangle =0.328$ GeV$^3$~\cite{Bodwin:2007fz}.

\begin{table}
{\centering
\begin{tabular}{@{}c@{}|@{}c@{}|@{}c@{}|@{}c@{}|@{}c@{}}
\hline 
 $\mu_r=\mu_f$& $\langle\mathcal{O}^{J/\psi}(^3S_1^{[8]})\rangle$ & $\langle\mathcal{O}^{J/\psi}(^1S_0^{[8]})\rangle$ & $\frac{\langle\mathcal{O}^{J/\psi}(^3P_0^{[8]})\rangle}{m_c^2}$ & $\, \, \frac{\chi^2_{\mathrm{min}}}{\text{d.o.f}}\, \, $ \\\hline 
\rowcolor{Myyellow}
$\, \, m_T/2\, \, $ & $\, \,0.592\pm 0.057\, \,$ & $\, \,-0.205\pm 0.196\, \,$   &   $\, \,0.697 \pm 0.089\, \,$ & $0.34$ \\\hline
\rowcolor{Myorange}
$\, \,  m_T\, \, $ &  $1.050\pm 0.121$ & $ 0.068\pm 0.2489$   &  $ 1.879 \pm 0.261$ & $ 0.22$ \\ \hline
\rowcolor{Myyellow}
$\, \, 2m_T\, \, $& $1.382\pm 0.189$ &$0.358\pm 0.303$    &  $3.270 \pm 0.533$ & $0.21$  \\ \hline
\end{tabular}
}
\caption{Fit results in units of $10^{-2}$ GeV$^3$ for our three scale choices $\mu_r=\mu_f=m_T/2$, $m_T$ and $2m_T$.}
\label{tab:LDMEfit}
\end{table}

\paragraph{Fit results ---} The best-fit values and the $\chi_\mathrm{min}^2$ per degree of freedom (d.o.f.) of our three fits are listed in Table \ref{tab:LDMEfit}. The respective correlation matrices are
\begin{align}
C_{m_T/2}&=
\begin{pmatrix}
 0.323 & -0.360 & 0.498\\
 -0.360 & 3.844 & -0.485 \\
 0.498 & -0.485& 0.789\\
\end{pmatrix}
\times 10^{-6}\, \text{GeV}^6,
\\
C_{m_T}&=
\begin{pmatrix}
 1.470 & -0.918 & 3.158\\
 -0.918 & 6.143 & -1.835 \\
 3.158 & -1.835& 6.838\\
\end{pmatrix}
\times 10^{-6}\, \text{GeV}^6,
\\
C_{2m_T}&=
\begin{pmatrix}
 3.577 & -1.895 & 10.053\\
 -1.895 & 9.157 & -5.077 \\
 10.053 & -5.077& 28.376\\
\end{pmatrix}
\times 10^{-6}\, \text{GeV}^6.
\end{align}

\begin{figure*}
\includegraphics[width=4.45cm]{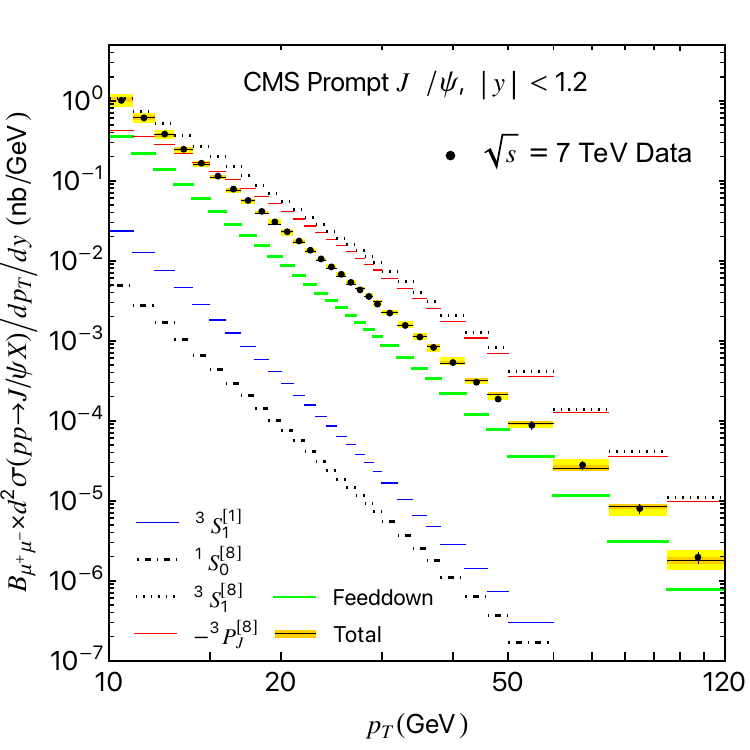}
\includegraphics[width=4.45cm]{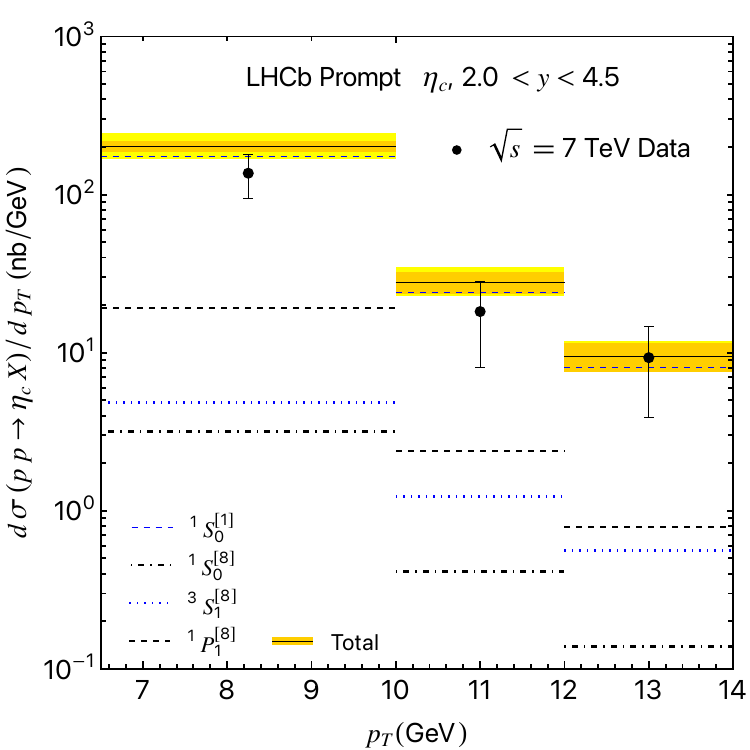}
\includegraphics[width=4.45cm]{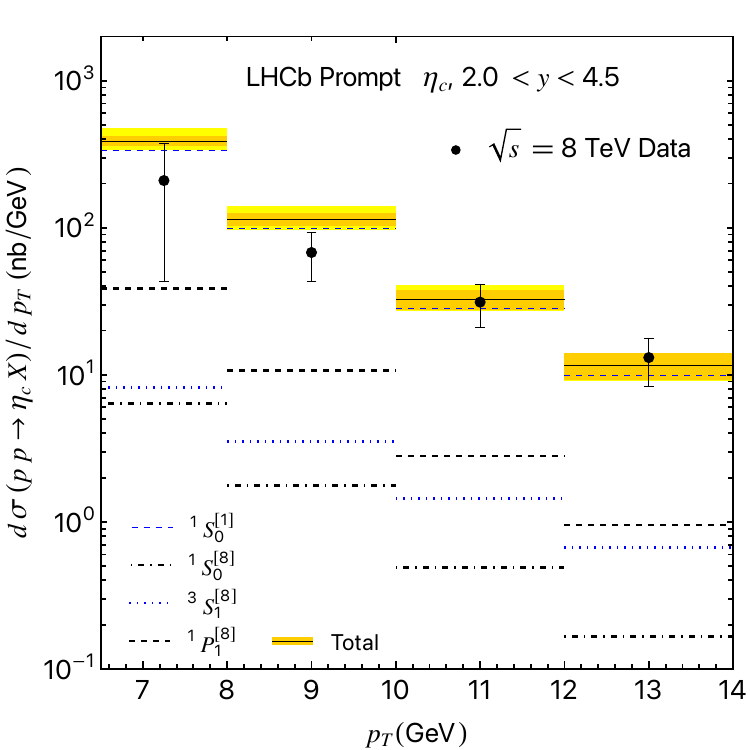}
\includegraphics[width=4.45cm]{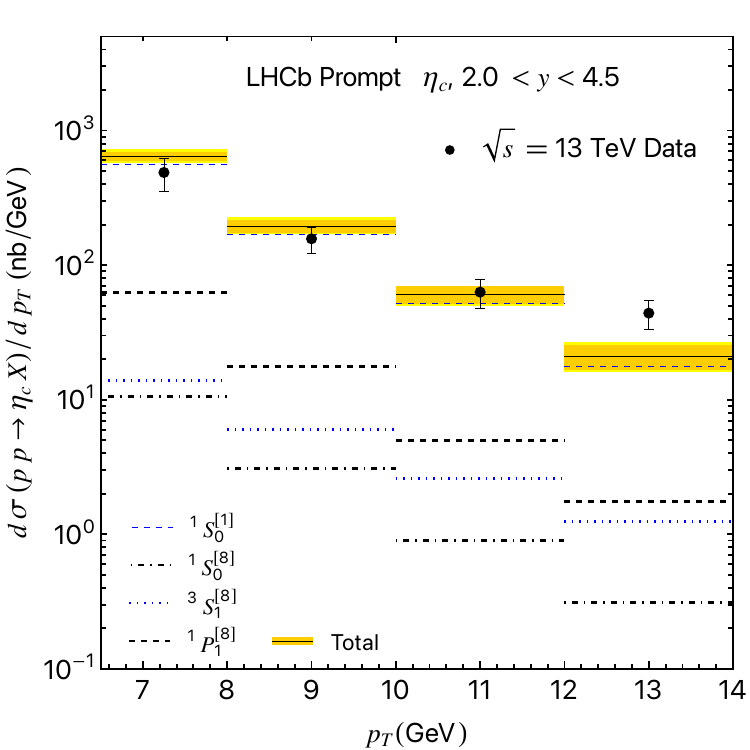}
\caption{\label{fig:fit}%
Predictions for prompt $J/\psi$ production at CMS~\cite{CMS:2015lbl} and $\eta_c$ production at LHCb~\cite{LHCb:2014oii,LHCb:2024ydi}. 
The data shown is exactly the data used for the LDME fits. 
The total cross section is broken down into feeddown contributions and the direct contributions of the individual Fock states. The orange bands describe the uncertainties from the fit correlations without considering scale variations. The yellow bands include the effect of both fit correlations and scale variations as described in the last paragraph of section {\em Introduction and Overview}.} 
\end{figure*}

In Fig.~\ref{fig:fit}, we compare the theoretical predictions using our fit results to the 42 data points. 
The central values are drawn by solid black lines and the uncertainties by colored bands, whereby the overall uncertainties, as described in the last paragraph of section {\em
Introduction and Overview}, are given by the yellow bands and the uncertainties only of the central fit by the orange bands inside them. 
We will use this convention for all plots throughout this publication. 
The level of agreement to the data apparent from Fig.~\ref{fig:fit} is in line with the very low values of $\chi^2_{\text{min}}/\text{d.o.f.}$ in Table~\ref{tab:LDMEfit}. 
The Fock state decomposition shows the familiar features~\cite{Han:2014jya,Zhang:2014ybe} of the high $p_T$ $J/\psi$ and $\eta_c$ fit: 
On the one hand, it is based on the cancellation between a large positive $^3S_1^{[8]}$ and a large negative $^3P_{J}^{[8]}$ $J/\psi$ production channel. 
We note here that this cancellation is not a fine-tuning problem, because NLO LDME mixing implies that only the sum of both contributions has physical significance, see e.g. section 3.2 of Ref.~\cite{Butenschoen:2019lef} for more details. 
On the other hand, $\langle{\cal O}^{J/\psi}({^1S}_0^{[8]})\rangle\approx\langle{\cal O}^{\eta_c}({^3S}_1^{[8]})\rangle$ is almost solely determined via $\eta_c$ production: 
With $\eta_c$ data already exhausted by the CS contribution, $\langle{\cal O}^{J/\psi}({^1S}_0^{[8]})\rangle$ is restricted to very small values.

\newcommand{\plotwithnumber}[2]{\begin{tikzpicture}
 \draw (0, 0) node[inner sep=0] {\includegraphics[width=4.45cm]{#1}};
 \draw (-1.69cm, -2.05) node {\tiny (#2)};
\end{tikzpicture}}

\begin{figure*}
\plotwithnumber{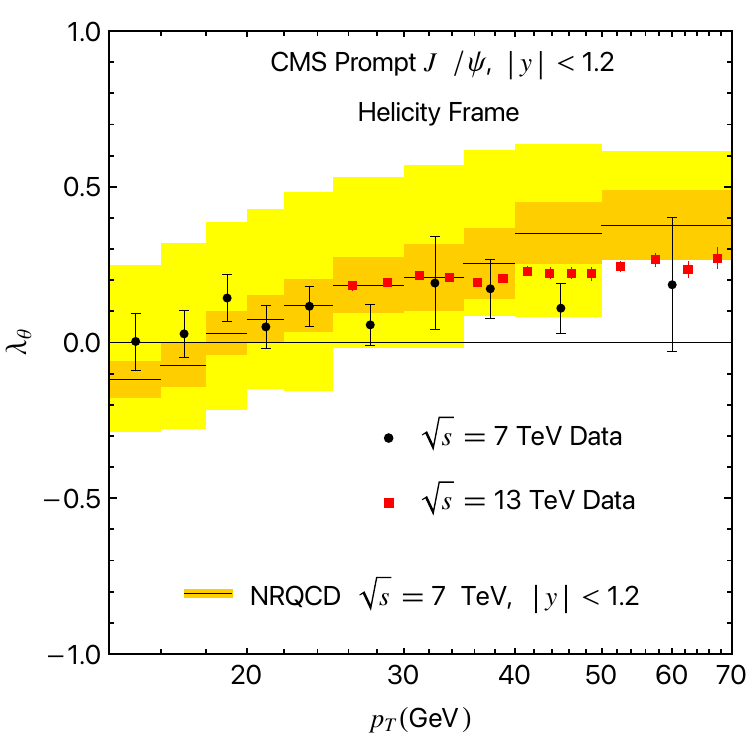}{a}
\plotwithnumber{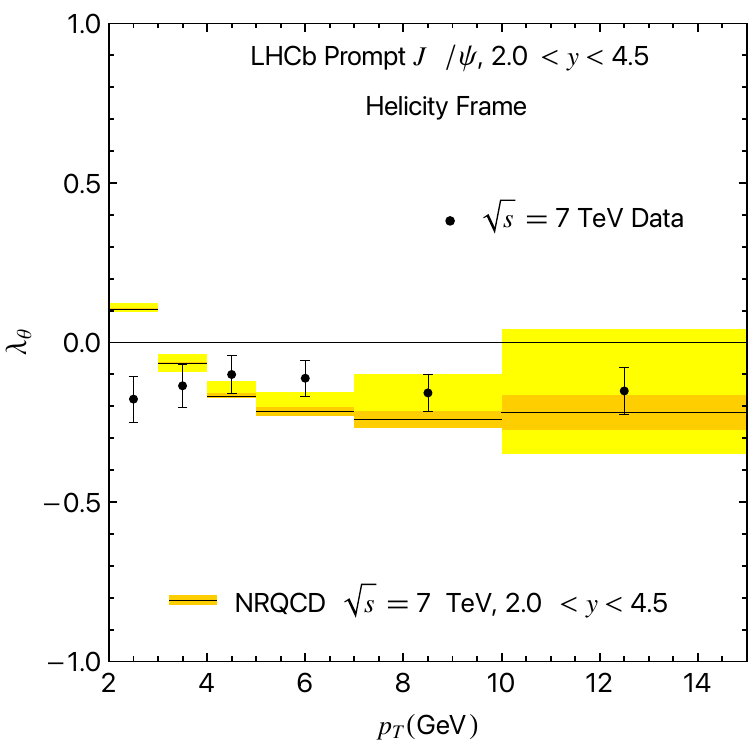}{b}
\plotwithnumber{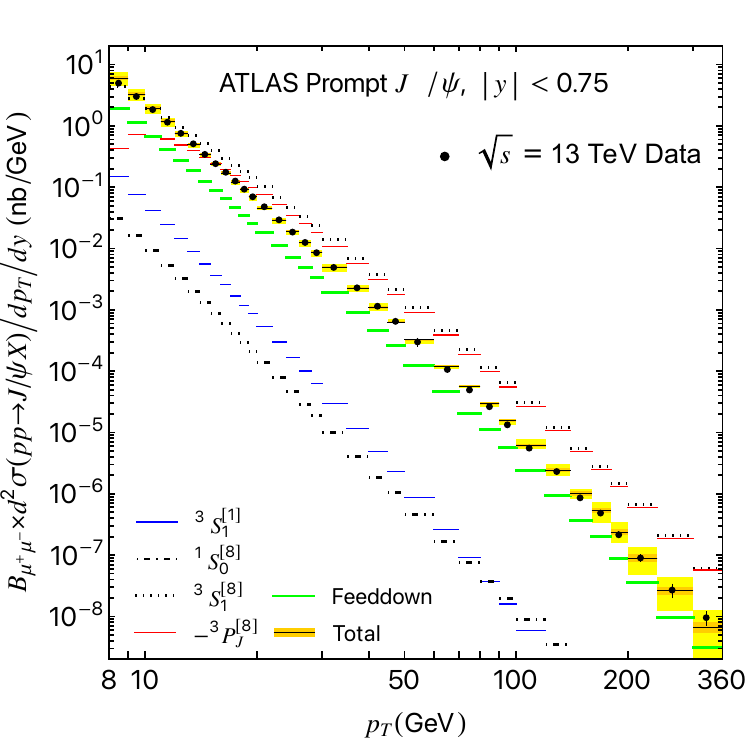}{c}
\plotwithnumber{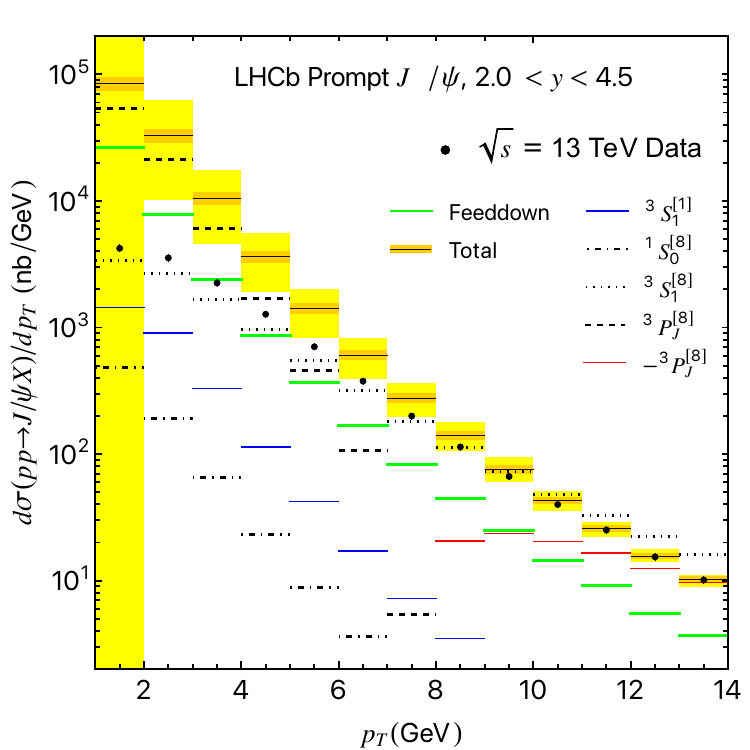}{d}
\plotwithnumber{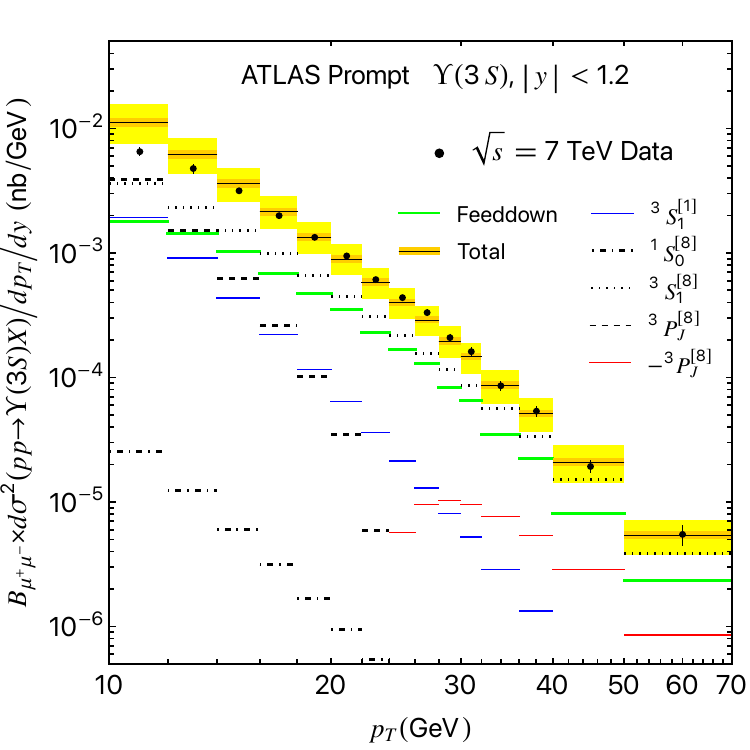}{e}
\plotwithnumber{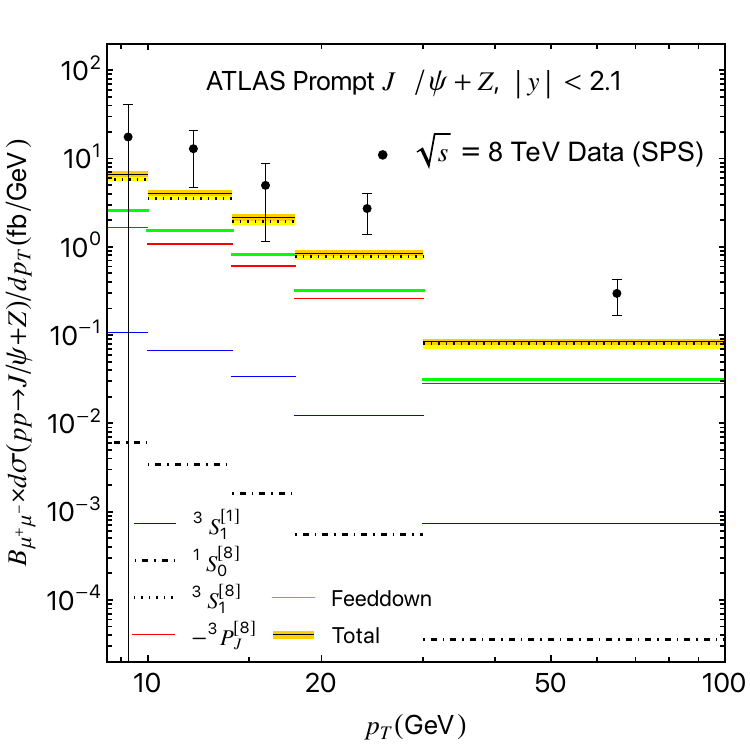}{f}
\plotwithnumber{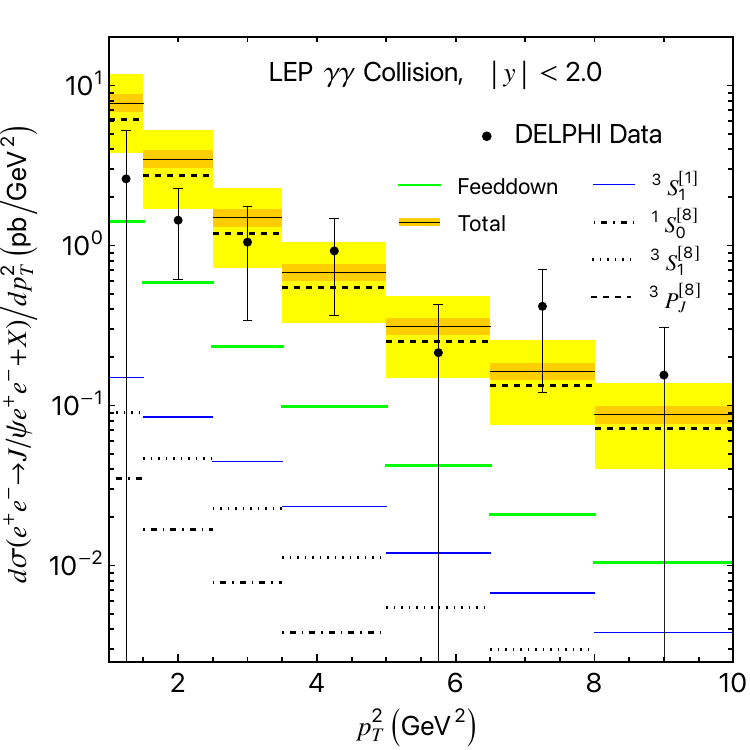}{g}
\plotwithnumber{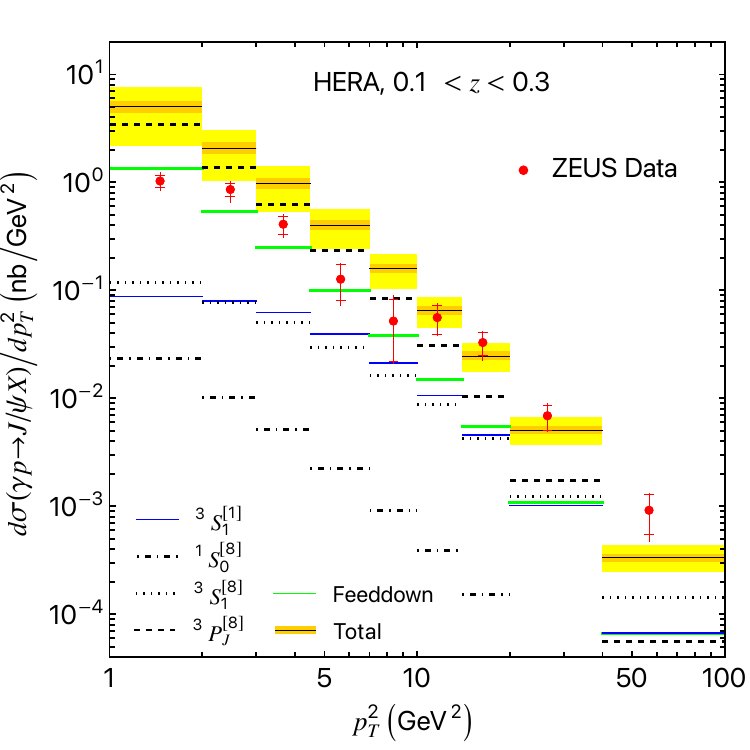}{h}
\plotwithnumber{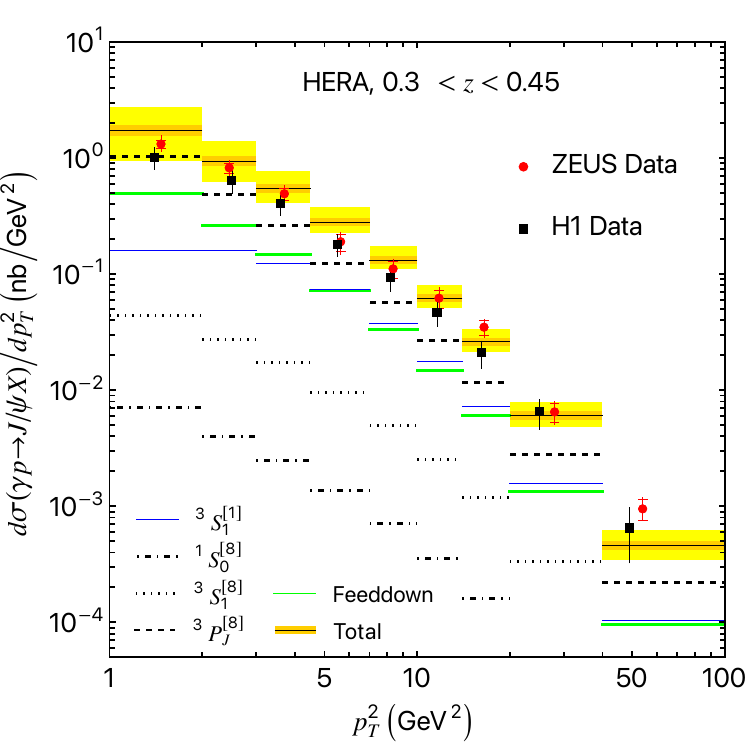}{i}
\plotwithnumber{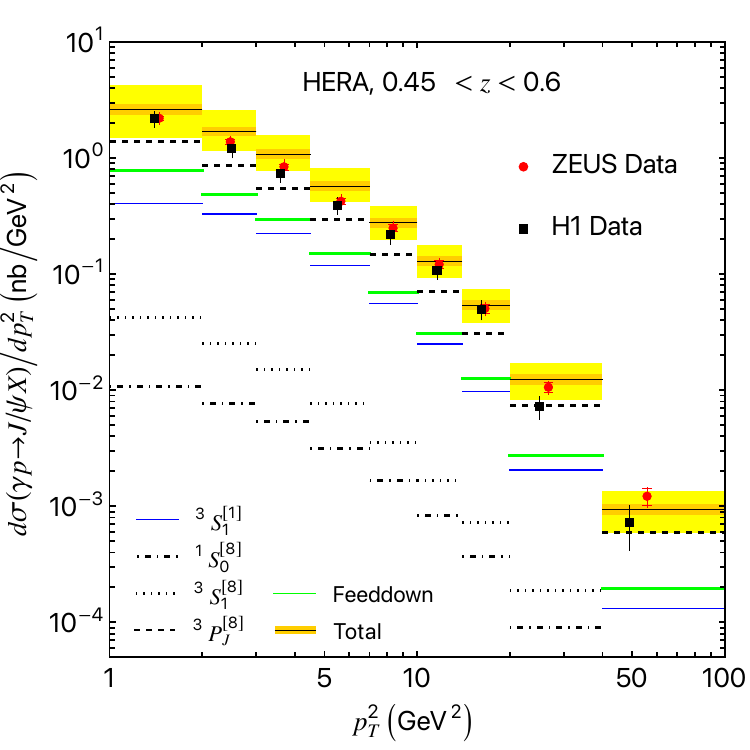}{j}
\plotwithnumber{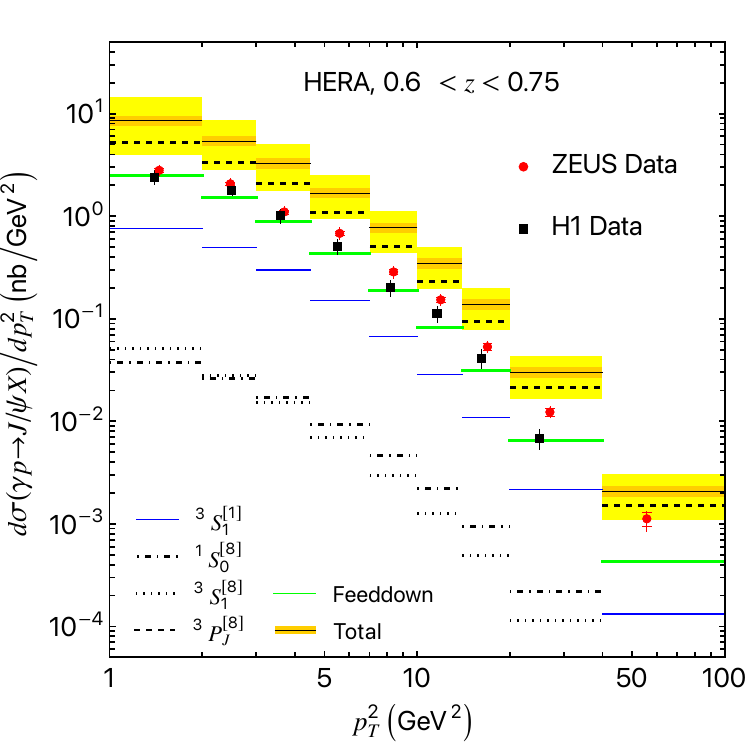}{k}
\plotwithnumber{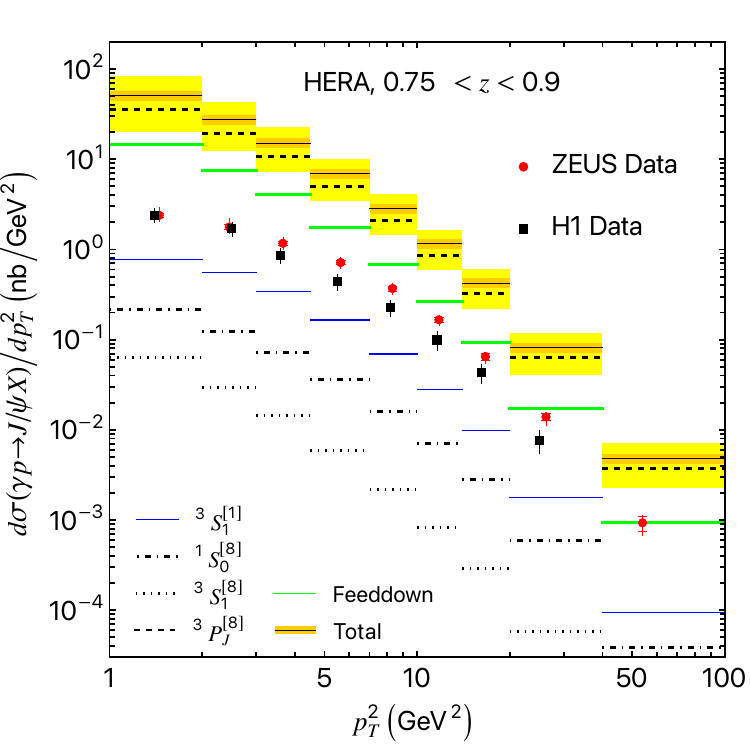}{l}
\caption{\label{fig:predictionplots}
Theory predictions for the $J/\psi$ polarization parameter $\lambda_{\theta}$ in the helicity frame compared to CMS~\cite{CMS:2013gbz,CMS:2024igk} (a) and LHCb~\cite{LHCb:2013izl} (b) data, for $J/\psi$ hadroproduction measured by ATLAS~\cite{ATLAS:2023qnh} (c) and LHCb~\cite{LHCb:2015foc} (d), for $\Upsilon(3S)$~\cite{ATLAS:2012lmu} (e) and $J/\psi+Z$~\cite{ATLAS:2014ofp} (f) production measured by ATLAS, for LEP DELPHI two photon collision data~\cite{DELPHI:2003hen} (g), and for HERA H1~\cite{H1:2010udv} and ZEUS~\cite{ZEUS:2012qog} photoproduction data (h--l).
For the $J/\psi+Z$ data shown in the plot, we have assumed $\sigma(pp\to Z+X)=(33.14\pm1.19)$~nb as in Ref.~\cite{Butenschoen:2022wld}, and subtracted the DPS contributions estimated in Ref.~\cite{ATLAS:2014ofp} using the DPS effective area value $\sigma_\mathrm{eff}=15^{+5.8}_{-4.2}$~mb of Ref.~\cite{ATLAS:2013aph}. To compare with the HERA $\gamma p$ data, we use ZEUS~\cite{ZEUS:2012qog} kinematics and divide the calculated $e^+e^-$ cross sections by the averaged photon flux $0.1024$~\cite{H1:2010udv}. Panels (g) to (l) show the sum of non-, single- (and double-) resolved photon contributions, which are not broken down here, to avoid clutter. In all panels, the total cross sections are broken down into feeddown contributions and the direct contributions of the individual Fock states. The orange bands describe the uncertainties from the fit correlations without considering scale variations. The yellow bands include the effect of both fit correlations and scale variations as described in the last paragraph of section {\em Introduction and Overview}.} 
\end{figure*}

\paragraph{Prediction of $J/\psi$ polarization ---}
Figs.~\ref{fig:predictionplots}a--\ref{fig:predictionplots}b show our predictions for the $J/\psi$ polarization parameter $\lambda_{\theta}$ in the helicity frame, compared to CMS~\cite{CMS:2013gbz,CMS:2024igk} and LHCb~\cite{LHCb:2013izl} data. 
Our predictions are in good agreement with these measurements and match the pattern that $\lambda_{\theta}$ turns from slightly negative at relatively low $p_T$  to positive and converges to $\lambda_{\theta}\sim 0.3$ at high $p_T$. 
 No $J/\psi$ polarization puzzle appears here.

\paragraph{Prediction of $J/\psi$ hadroproduction at very high $p_T$---}
The recent ATLAS measurement~\cite{ATLAS:2023qnh} of the $J/\psi$ $p_T$ differential distribution at $\sqrt{s}=13$ TeV with $p_T$ ranging up to $360$ GeV, which we compare our predictions to in Fig.~\ref{fig:predictionplots}(c), has received significant attention. 
In the publication itself, data was compared to NLO NRQCD predictions using LDME input from Ref.~\cite{Butenschoen:2011yh}. 
Although there were no unreasonable discrepancies, theory errors were expectedly large at these very high $p_T$ values with their dominating $\log(m_c^2/p_T^2)$ terms. 
This effect depends however on the LDME set chosen. We can fortunately infer a negligible effect on our results from Fig.~3 of Ref.~\cite{Chung:2024jfk}, where the {\em resummed} curve of the {\em Chao et al.~\cite{Han:2014jya} LDMEs} panel describes the ATLAS data~\cite{ATLAS:2023qnh} well, noticing that the LDMEs of Ref.~\cite{Han:2014jya} do not significantly differ from our default LDMEs.

\paragraph{Prediction of $J/\psi$ hadroproduction at low $p_T$ ---} In Fig.~\ref{fig:predictionplots}d, we compare our predictions to $J/\psi$ production data from LHCb~\cite{LHCb:2015foc}, reaching down to $p_T=1$~GeV. 
Here, we recover the feature familiar from other high $p_T$ fits~\cite{Ma:2010yw,Gong:2012ug,Han:2014jya,Zhang:2014ybe,Bodwin:2015iua,Brambilla:2022rjd,Brambilla:2022ayc} that low $p_T$ data is not reproduced. 
The  reason is that the $^3P_{J}^{[8]}$ SDCs change sign from negative to positive when going below $p_T\approx 7$~GeV, 
so that instead of a cancellation between $^3S_1^{[8]}$ and $^3P_{J}^{[8]}$ channels there is an amplification. 
The resulting steep increase at low $p_T$ is not observed in the data.

\paragraph{Prediction of $\Upsilon(3S)$ hadroproduction ---}
Fig.~\ref{fig:predictionplots}e shows our predictions for the $p_T$ differential~$\Upsilon(3S)$ production measured by ATLAS~\cite{ATLAS:2012lmu}. 
These results are a genuine prediction of the equations~(3.47)--(3.48) of Ref.~\cite{Brambilla:2022ayc} derived from pNRQCD, 
which relate $\Upsilon(3S)$ to the $J/\psi$ CO LDMEs. 
The accurate description seen in Fig.~\ref{fig:predictionplots}e can thus also be viewed as a highly nontrivial confirmation of the pNRQCD relations, 
the more so as scale evolutions in equation~(3.47) of Ref.~\cite{Brambilla:2022ayc}
result in a very different Fock state decomposition in $\Upsilon(3S)$ as compared to $J/\psi$ production. 
In particular, there is no strong cancellation between $^3S_1^{[8]}$ and $^3P_{J}^{[8]}$ channels here. Instead, the cross section decomposes almost equally into the $^3S_1^{[8]}$ channel and feeddown from $\chi_{bJ}$ mesons. 
For $\Upsilon(1S)$ and $\Upsilon(2S)$, we reach similar conclusions.

\paragraph{Prediction of LHCb $J/\psi+Z$ production ---} In Fig.~\ref{fig:predictionplots}f, we show our predictions for $J/\psi+Z$ production measured by ATLAS~\cite{ATLAS:2014ofp}, 
where estimated double parton scattering (DPS) contributions are already subtracted from the data. 
The predictions are dominated by the $^3S_1^{[8]}$ channel. 
For the two highest $p_T$ bins, predictions lie around two experimental standard deviations below data. 
Given the fact that in Ref.~\cite{ATLAS:2014ofp}, a data-driven analysis of the $J/\psi$-$Z$ angular distribution confirmed single parton scattering (SPS) dominance, it appears unlikely, albeit not impossible, 
that the discrepancy is due to underestimated DPS contributions. This conclusion is however based on the assumption that the pocket formula underlying the ATLAS DPS estimation holds, which itself is, however, subject to debate.

\paragraph{Prediction of $J/\psi$ production in $\gamma\gamma$ scattering ---}
In Fig.~\ref{fig:predictionplots}g, we show predictions for $p_T^2$ differential $J/\psi$ production in $\gamma\gamma$ scattering at LEP DELPHI~\cite{DELPHI:2003hen}, in the range 1~GeV$^2<p_T^2<10$~GeV$^2$. 
Contrary to the hadroproduction case, this low $p_T$ data is reasonably well described, even though the very low statistics of the measurement does not allow for more than an order-of-magnitude comparison. 
We note that the cross sections are almost exclusively given by single-resolved photon contributions.

\paragraph{Predictions of $J/\psi$ production in HERA photoproduction ---} In Figs.~\ref{fig:predictionplots}h--\ref{fig:predictionplots}l, we show our predictions for $p_T^2$ differential $J/\psi$ photoproduction at HERA~\cite{H1:2010udv,ZEUS:2012qog} for five bins of the inelasticity variable $z$, which measures in the proton rest frame the fraction of photon energy transferred to the $J/\psi$. 
In the region $0.1<z<0.3$ with $p_T^2<40$~GeV$^2$, resolved photons are dominating our predictions, while for the other regions, nonresolved photons are dominating.
We recover the increase of the cross section at $z\to 1$, known e.g. from Fig.~3.3 of Ref.~\cite{Boer:2024ylx}, which is not observed in the data. 
For $0.75<z<0.9$, predictions overshoot the data by factors of 5.2 to 20. 
Surprisingly, in the $z<0.6$ regions, we obtain a good description of the data over the whole measured $p_T$ range, 
again down to $p_T=1$~GeV, with a significant deviation only for the low $p_T$ end of the lowest $z$ bin, which coincides with the region where resolved photoproduction dominates. 

\paragraph{Discussion ---} Although NLO predictions of the high $p_T$ $J/\psi$ plus $\eta_c$ combined fit presented here lead to many nontrivial descriptions of various quarkonium production data, discrepancies remain. 
Moreover, these discrepancies are not at places where one would naively expect them. 
A naive expectation would be that fixed order NLO calculations may fail at $p_T^2\gg 4 m_Q^2$ and in the endpoint regions $p_T^2\ll 4 m_Q^2$ and $1-z\ll 1$, 
where large logarithms spoil the perturbative convergence. 
The hadroproduction predictions from the LDME fit presented here are, however, very good for $p_T^2\gg 4 m_Q^2$, 
and fail only in the region $p_T^2\lessapprox 4 m_Q^2$. 
On the other hand, the latter region works well for $z<0.6$ HERA photoproduction, 
while for $z>0.6$, HERA data is not described, regardless of $p_T$. 
LDME universality thus remains in question. 
However, as pointed out above, we could find indications from Ref.~\cite{Chung:2024jfk} that, for our LDMEs, resummations in the large $p_T$ hadroproduction limit would yield no large effect. Furthermore, it is noticeable that the remaining regions of disagreement, $p_T^2\lessapprox 4m_Q^2$ in hadroproduction and $z>0.6$ in photoproduction, coincide with extensions of endpoint regions, for which solutions, e.g. via small-$x$ resummation \cite{Ma:2014mri} and nonperturbative shape functions \cite{Beneke:1997qw,Fleming:2006cd}, have been proposed. This is also true for the BELLE  cross section $e^+e^-\rightarrow J/\psi + X_{{\rm non}~c\bar{c}}$~\cite{Belle:2009bxr}, see Refs.~\cite{Beneke:1997qw,Chen:2022qli},  where we have tested that using the SDCs from Ref.~\cite{Chen:2022qli} and our default fit LDMEs, we obtain a good description of the BELLE data.

\paragraph{Summary ---} To summarize, we have done a combined fit of the three $J/\psi$ CO LDMEs to 42 data points of CMS $J/\psi$ and LHCb $\eta_c$ production data and made predictions for CMS and LHCb $J/\psi$ polarization, ATLAS high $p_T$ $J/\psi$ production, ATLAS $\Upsilon(3S)$ and $J/\psi+Z$ production, LHCb low $p_T$ $J/\psi$ production as well as $J/\psi$ production in $\gamma \gamma$ collisions at LEP and $\gamma p$ collisions at HERA. 
We have thereby used a novel fit-and-predict procedure, which systematically takes into account the effect of scale variations. 
The predictions show good agreement with data everywhere with the exception of low and medium $p_T$ hadroproduction, HERA photoproduction at $z>0.6$, and possibly the two highest LHCb $J/\psi+Z$ production bins. 
In particular, our findings for $J/\psi$ production in $\gamma \gamma$ and $\gamma p$ scattering may be of significance for future quarkonium studies, in particular at the EIC and the high-luminosity LHC.

\begin{acknowledgments}
X.-P.~W would like to thank A. Bruni, Y.-Q. Ma and H.-S. Shao  for useful discussions. We thank A. Vairo for reading the paper and giving very useful comments.
The work of N.~B. and X.-P.~W. is supported by the DFG (Deutsche Forschungsgemeinschaft,
German Research Foundation) Grant No. BR 4058/2-2. 
We acknowledge support from the DFG cluster of excellence ``ORIGINS'' under
Germany's Excellence Strategy - EXC-2094 - 390783311. X.-P.~W. acknowledges support from STRONG-2020- European Union’s Horizon 2020 research and innovation program under grant agreement
No. 824093. N. B. acknowledges the European Union ERC-2023-ADG- Project EFT-XYZ. The work of X.-P.~W.  is supported by the National Natural Science Foundation of China under Grant No.~12135006.
The work of M.~B. is supported by the German Research Foundation DFG through Grant No. BU 3455/1-1 as part of the Research Unit FOR292. 
The authors would like to express special thanks to the Mainz Institute for Theoretical Physics (MITP) of the Cluster of Excellence PRISMA+ (Project ID 390831469), for its hospitality and support.

\end{acknowledgments}

\bibliography{NLOproduction.bib}

\end{document}